\documentclass[prd,amsmath,preprintnumbers,superscriptaddress,floatfix,nofootinbib,letterpaper,english]{revtex4}
\usepackage{amssymb}
\usepackage{amsmath}
\usepackage{amsfonts}
\usepackage{epsfig}
\usepackage{graphicx}
\usepackage{tabularx}
\usepackage{latexsym}
\usepackage{subfigure}
\usepackage{verbatim}
\usepackage{color}
\usepackage{braket}
\usepackage{multirow}
\usepackage{dcolumn}
\usepackage{mathtools}
\usepackage{placeins}
\usepackage{epstopdf}
\usepackage{wrapfig}
\usepackage{hyperref}
\usepackage{diagbox}
\usepackage{booktabs}
\usepackage{simplewick}
\usepackage[normalem]{ulem}
\usepackage{extarrows}
\usepackage{lipsum,babel}
\usepackage[utf8]{inputenc} 
\usepackage[toc,page]{appendix}
\usepackage{mathtools}
\def\gev{\mathrm{GeV}}

\renewcommand\sout{\bgroup \color{red} \ULdepth=-.5ex \ULset}

\newcommand{\sub}[1]{\begin{subequations}
			#1
		     	     \end{subequations}}

\DeclarePairedDelimiter\abs{\lvert}{\rvert}%
\allowdisplaybreaks[1] 
\newcommand{\bm}[1] {\mbox{\boldmath{$#1$}}}

\def\eq{\begin{eqnarray}}
\def\en{\end{eqnarray}}
\def\ep{\epsilon}
\newcommand{\la}{\langle}
\newcommand{\ra}{\rangle}
\def\nonu{\nonumber \\ &&}
\usepackage{footnote}

\begin{document}
\preprint{}


\title{Gravitational form factors of $\rho$ meson with a light-cone constituent quark model}

\author{Bao-Dong Sun}
\email{sunbd@sdu.edu.cn}
\affiliation{Key Laboratory of Particle Physics and Particle Irradiation, Ministry of Education,
Institute of Frontier and Interdisciplinary Science, Shandong University,
Shandong 266237, China}
\affiliation{Institute of High Energy Physics, Chinese Academy of Sciences, Beijing 100049, People's Republic of China}
\affiliation{School of Physics, University of Chinese Academy of Sciences, Beijing 100049, People's Republic of China}
\author{Yu-Bing Dong}
\email{dongyb@ihep.ac.cn}
\affiliation{Institute of High Energy Physics, Chinese Academy of Sciences, Beijing 100049, People's Republic of China}
\affiliation{School of Physics,  University of Chinese Academy of Sciences, Beijing 100049, People's Republic of China}
\affiliation{Theoretical Physics Center for Science Facilities (TPCSF), CAS, Beijing 100049, People's Republic of China}
\noaffiliation

\begin{abstract}
The $\rho$ meson gravitational form factors are studied based on a light-front constituent quark model
which has been successfully employed to calculate its generalized parton
distributions and some low-energy observables.  The distributions of energy, spin, pressures, and shear
forces inside the $\rho$ meson are explicitly given.
\end{abstract}
\date{\today}
\maketitle

\section{Introduction}
\label{intro}

We know that the gravitational form factors (GFFs) are defined through the matrix element of the energy-momentum tensor
(EMT)~\cite{Pagels:1966zza}. Since the GFFs relate to the mass, spin, shear forces, and $D$-term of the
particles~\cite{Goeke:2001tz,Polyakov:2002yz}, they (or EMT form factors) involve a large range of physics, such as the gravitation
physics and the physics in hard scattering processes~\cite{Polyakov:2018zvc,Lorce:2018egm}.  It is a promising way to extract more
information about the mechanical properties of a hadron (especially in the non-perturbative region) from the study of
GFFs. Those tasks mainly try to answer some fundamental questions, like how the hadron mass and spin are carried out
by quarks and gluons or what the mechanism that the trace anomaly contributes to hadron mass is, and how the strong
force distributes inside the hadron, {\it etc.}. Besides, the Fourier transforms of the EMT matrix elements define
the static EMT which can further tell the distributions of pressure and shear forces~\cite{Polyakov:2018zvc,Lorce:2018egm}.\\

In the 1960s, the total GFFs were already introduced for both spin-0 and spin-1/2 hadrons~\cite{Pagels:1966zza}.
The most natural but also the least practical way to probe GFFs is scattering processes through graviton
exchange. However, it's more
practical to extract GFFs through their connections to the generalized parton distributions (GPDs). The relations
between GFFs and GPDs were discussed in detail in Refs.~\cite{Ji:1996ek,Polyakov:1999gs}. As the soft part of the
hard-exclusive reactions, GPDs have been received many theoretical and experimental
investigations~\cite{Diehl:2003ny, Belitsky:2005qn,Ji:2006ea,Frederico:2009fk,Diehl:2015uka,Pasquini:2019evu}.
As a reflection of the broken scale invariance of QCD, the matrix element of the trace anomaly part of EMT naturally
connects with the hadron mass~\cite{Ji:1994av,Yang:2018nqn}. Especially the gluonic operator is believed to contribute
to the majority part. This may give another possible way to probe the GFFs via the exclusive production of heavy
quarkonium states, such as near-threshold $J/\psi$ and $\Upsilon$ photoproduction processes at JLab. and RHIC
{\it etc.}~\cite{Kharzeev:1995ij,Kharzeev:1998bz,Hatta:2019lxo}. \\

At present, it is still not clear what the specific relations among the strong force, pressure and shear forces
are. Nevertheless, one may get some hints from the phenomenological studies of the static EMT of particles with
different spins. For instance, the GFFs of pion (spin-0) were evaluated in chiral quark models in
Refs.~\cite{Broniowski:2007si,Broniowski:2008hx} and parameter methods \cite{Kumano:2017lhr}. Ref.~\cite{Hudson:2017xug} applies the Q-ball model to spin-0
particles as well, where the $D$-term, energy density, pressure, and shear forces were investigated in detail. There
are also Lattice QCD calculations related to the pion GFFs \cite{Best:1997qp}. For the spin-1/2 hadrons, there are
model calculations from the AdS/QCD approach~\cite{Anikin:2019ufr}, and the chiral quark soliton
model~\cite{Goeke:2007fp}, {\it etc.}. More can be found in a review article (see Ref.~\cite{Polyakov:2018zvc}). The formalism of GFFs for a spin-1 hadron are discussed by
Refs.~\cite{Holstein:2006ud,Polyakov:2019lbq,Cosyn:2019aio} and for arbitrary spin hadrons in recent Ref.~\cite{Cotogno:2019vjb}. In the literature, the model calculations for the spin-1 particles
include the AdS/QCD approach~\cite{Abidin:2008ku} and the Nambu-Jona-Lasinio (NJL) model~\cite{Freese:2019bhb}. \\

It is shown that the light-cone quark model (LCCQM) for the $\rho$ meson employed in our previous works can describe
the $\rho$ meson well in the low energy region, such as its electromagnetic form factors, GPDs, {\it etc.}~
\cite{Sun:2017gtz,Sun:2018tmk,Sun:2018ldr}. In this work, we'll apply our LCCQM and the previous results of GPDs to study
the $\rho$ meson GFFs and its mechanical properties (quadrupole pressure and shear forces, {\it etc.}~\cite{Polyakov:2019lbq,Panteleeva:2020ejw}), and try to get some information about those fundamental questions.\\

This paper is organized as follows. In Section~\ref{sec:1}, the definitions of GFFs, pressure, and shear forces for a spin-1
particle are briefly presented. Moreover, the LCCQM for the $\rho$ meson applied in our previous works is also
shortly reviewed in this section. Section~\ref{sec:Results} gives our numerical results for the $\rho$ meson GFFs,
pressure, and shear forces, {\it etc.}. In this part, we introduce a phenomenological three-dimensional (3D) Gaussian
form wave package, when calculating the static EMT, since our obtained GFFs do not drop fast enough. We also display
our model-dependent $D$-term of the $\rho$ meson, which is not affected by the Gaussian form wave package.
Finally, section~\ref{sec:Summary} is devoted for a summary.\\

\section{GFFs of spin one particles and our model}
\label{sec:1}

\subsection{GFFs of spin one particles}
\label{sec:gffs}

The formalism of GFFs of a spin one particle and its other mechanical properties have been discussed and given
explicitly~\cite{Holstein:2006ud,Polyakov:2019lbq,Cosyn:2019aio}. Here, we briefly summarize them as follows.
In this paper, we use the covariant normalisation $\la p^\prime, \sigma^\prime|\,p,\sigma\ra=
2p^0\,(2\pi)^3\delta^{(3)}({\vec p}^{\,\prime}-\vec{p}\,)\delta_{\sigma\sigma^\prime}$ for the system, and introduce the
kinematic variables $P= \frac12(p^\prime + p)$, $\Delta = p^\prime-p$, $t=\Delta^2$. Then, the symmetric (Belinfante) EMT form factors of a
spin-1 particle in QCD are defined as,
\eq  \label{Eq:EMT-FFs-spin-1}
\langle p^\prime,\sigma^\prime| \hat T_{\mu\nu}^a(x) |p,\sigma\rangle
&=& \biggl[
2P_\mu P_\nu  \Bigl(
- {\ep^{\prime*}\cdot \ep} \, A^a_0 (t)
 +{ {\ep^{\prime*}\cdot P} \, {\ep \cdot P} \over m^2}
 \, A^a_1(t) \Bigl)
\nonumber\\
&&+2\left[P_\mu(\ep^{\prime*}_{\nu} \,\ep\cdot P+\ep_{\nu}\,
\ep^{\prime*}\cdot P)
+P_\nu(\ep^{\prime*}_{\mu}\, \ep\cdot
P+\ep_{\mu} \,\ep^{\prime*}\cdot
P) \right] \, J^a (t)
\nonumber\\
&&+\frac12(\Delta_\mu \Delta_\nu-g_{\mu\nu}\Delta^2)
 \Bigl(
{\ep^{\prime*}\cdot \ep} \, D^a_0 (t)
+{ {\ep^{\prime*}\cdot P} \, {\ep \cdot P} \over m^2}  \, D^a_1(t)\Bigl)
\nonumber\\
&&+\Bigl[\frac12(\ep_{\mu}
\ep^{\prime*}_{\nu}+\ep^{\prime*}_{\mu}\ep_{\nu})\Delta^2
-(\ep^{\prime*}_{\mu}\Delta_\nu+\ep^{\prime*}_{\nu} \Delta_\mu)\,\ep\cdot P
\nonumber\\
&&
+(\ep_{\mu} \Delta_\nu+\ep_{\nu}
\Delta_\mu)\,\ep^{\prime*}\cdot P
-4g_{\mu\nu} \, {\ep^{\prime*}\cdot P} \, {\ep\cdot P} \Bigl] \, E^a(t)
\nonumber\\
&&
+\Bigl(\ep_{\mu}
\ep^{\prime*}_{\nu}+\ep^{\prime*}_{\mu}\ep_{\nu} - \frac{{\ep^{\prime*}\cdot \ep}}{2}\,g_{\mu\nu} \Bigl)
\,{m^2} \, {\bar f}^a (t) \nonumber\\
&&+g_{\mu\nu} \Bigl( {\ep^{\prime*}\cdot \ep} \, {m^2}\, {\bar c}^a_0(t)\, +  \, {\ep^{\prime*}\cdot P} \, {\ep \cdot P}
\,{\bar c}^a_1(t)  \Bigl)  \biggr] \,e^{i(p^\prime-p)x} \ ,
\en
where $m$ is $\rho$ meson mass and $a=g,u,d,\dots$, which represent the contributions of gluon and all flavors of quarks, and the polarization
vectors $\ep^\prime_\mu=\ep_\mu(p^\prime,{\sigma^\prime})$, $\ep_{\mu}=\ep_{\mu}(p,\sigma)$
with $\sigma=x,y,z$, respectively. The 6 quark and gluon GFFs $A^a_{0,1}$, $D^a_{0,1}$, $J^a$ and  $E^a(t)$ are
individually momentum-energy conserving, and the other 3 GFFs, ${\bar f}^a$ and ${\bar c}^a_{0,1}(t)$, are not. \\

As shown in our previous works, in the Breit frame, the above expression can be re-organized according to the power of the
quadrupole operator of the spin one particles. The static EMT $T^{\mu\nu}(\vec r, \sigma^\prime,\sigma)$
of the spin-1 system is defined by the Fourier transform of the EMT with respect to $\vec \Delta$ as
\eq \label{eq:static_EMT}
T^{\mu\nu}_a(\vec r, \sigma^\prime,\sigma) &=& \int {d^3 \Delta \over 2E(2\pi)^3 } e^{-i \vec \Delta \cdot \vec r}
\langle p^\prime, \sigma^\prime \, |{\hat T}^{\mu\nu}_a(0)|p,\sigma \rangle \ .
\en
Eq. \ref{eq:static_EMT} contains the energy densities, the distributions of spin, pressure and shear forces with different
power of quadrupole operator. For the energy distributions, we have (sum over all gluons and quark flavors)
\eq
T^{00} ( \vec r, \sigma^\prime,\sigma)&=&
\int {d^3 \Delta \over 2E (2\pi)^3 } e^{-i \vec \Delta \cdot \vec r}
\langle p^\prime, \sigma^\prime \, |{\hat T}^{00}(0)|p,\sigma \rangle
\\
&=& \label{eq:thetauv00BA}
\varepsilon_0(r) \, \delta_{\sigma^\prime\sigma}
+\varepsilon_2(r)\, {\hat Q}^{ij} \, Y^{ij}_2
\ ,
\en
where $r = \abs{\vec r \,}$, $Y^{ij}_2=r^i r^j/r^2-\delta^{ij}/3$, ${\hat Q}^{ij}=({\hat Q}^{ij})_{\sigma'\sigma}$, and
\sub{
\eq
\varepsilon_0(r) &=& m\,{\tilde {\mathcal E}}_{0}(r) \ , \\
~~~~~\varepsilon_2(r) &=& -\frac{1}{2m}{r} {d\over dr} {1\over r} {d\over dr} {\tilde {\mathcal E}}_{2}(r) \ , \
\label{eq:ft2E}
\en }
with
\eq
\quad {\tilde {\mathcal E}}_{0,2}(r) =2m\int{d^3\Delta\over 2E(2\pi)^3} e^{-i\vec\Delta\cdot\vec r} {\mathcal E}_{0,2}(t) \,
\en
where ${\mathcal E}_{0,2}(t)=\sum_a {\mathcal E}_{0,2}^a(t)$ and, in the Breit frame,
$t=-\vec\Delta^2$ and $E=\sqrt{m^2+\vec\Delta^2/4}$. \\

For the spin distribution, the $0j$ component is
\eq \label{Eq:EMT-Breit-T0k}
T^{0j}_a (\vec r, \sigma^\prime,\sigma)
&=&
\int {d^3 \Delta \over 2E (2\pi)^3 } e^{-i \vec \Delta \cdot \vec r}
\langle p^\prime, \sigma^\prime \, |{\hat T}_a^{0j}(0)|p,\sigma \rangle \ .
\en
The individual contributions of quarks and gluons to the spin of the particle is
\eq \label{Eq:static-EMT-T0k-2}
	J_a^i(\vec{r},\sigma^\prime,\sigma)
	&=& \epsilon^{ijk}r^jT^{0k}_a(\vec{r},\sigma^\prime,\sigma)\;, \\
&=&{\hat S}^j_{\sigma^\prime\sigma}  \int {d^3 \Delta \over  (2\pi)^3 } e^{-i \vec \Delta \cdot \vec r}
\bigg[ \Big(  \mathcal{\bar J}^a(t) +\frac23 t {d \mathcal{\bar  J}^a(t) \over dt}  \Big) \delta^{ij}
+ \Big( \Delta^i\Delta^j- \frac13 \vec \Delta^2 \delta^{ij}  \Big) {d \mathcal{\bar J}^a(t) \over dt}  \bigg] \; ,
\en
with $\mathcal{\bar J}^a(t)=\frac{m}{E} \mathcal{J}^a(t)$,
and the spin operator~\cite{Varshalovich:1988ye,Polyakov:2019lbq}\footnote{In Ref.~\cite{Polyakov:2019lbq}, the spin operator and rest frame spin-1 polarization vectors in Eq.(13) and (14) are incorrect.}.
\eq
\hat S^{\ i}_{\sigma^\prime \sigma}= -i \ep^{ijk} \, \ep_{\sigma}^{*\,j}  \ep_{\sigma^\prime}^k \ , \ (i,j,k,\sigma^\prime,\sigma=x,y,z),
\en
where the rest frame spin-1 polarization vectors are
\eq
 \ep_x =
\begin{pmatrix}
1\\0\\0
\end{pmatrix}
\ , \quad
 \ep_y =
\begin{pmatrix}
0\\1\\0
\end{pmatrix}
\ , \quad
 \ep_z =
\begin{pmatrix}
0\\0\\1
\end{pmatrix}
\ .
\en \\

For the $ij$-components, the quadrupole elastic pressure and shear forces are firstly defined in Ref.~\cite{Polyakov:2019lbq} in the sprit of Ref.~\cite{Lorce:2017wkb,Polyakov:2018rew,Schweitzer:2019kkd}. A new parameterization of pressure and shear forces is introduced in a recent paper~\cite{Panteleeva:2020ejw}, and it conveniently generates the normal and tangential forces acting on radial area element ($dF_r$, $dF_\theta$ and $dF_\phi$). These forces explains how the hadron shape forms. The corresponding relations of two sets of parameterization are given in Appendix of Ref.~\cite{Panteleeva:2020ejw}.  According to Ref.~\cite{Panteleeva:2020ejw}, 
\eq  \label{eq:thetauvij_2}
T^{ij}({\vec r}) &=& \int {d^3 \Delta \over 2E(2\pi)^3 } e^{-i \vec \Delta \cdot \vec r}
\langle p^\prime, \sigma^\prime \, |{\hat T}^{ij}(0)|p,\sigma \rangle  \\
& =&p_0(r) \delta^{ij}+s_0(r)Y_2^{ij} + \left(p_2(r)+\frac13 p_3(r)-\frac19 s_3(r)\right) \hat{Q}^{ij}   \\
\label{eq:quadrupoleik}
&&+ \left( s_2(r)-\frac12 p_3(r)+\frac16 s_3(r) \right) 
2 \left[\hat{Q}^{ip}Y_{2}^{pj}+\hat Q^{jp}Y_{2}^{pi} -\delta^{ij} \hat Q^{pq}Y_{2}^{pq}    \right]  \nonumber \\
&&+ \hat Q^{pq}Y_{2}^{pq}\left[\left(\frac23 p_3(r)+\frac19 s_3(r)\right)\delta^{ij}+\left(\frac12 p_3(r)+\frac56 s_3(r)\right) Y_2^{ij}\right]  +\ldots \nonumber
\en
where the quadrupole pressure $ p_n(r)$ and shear forces functions  $ s_n(r)$ can be written as
\sub{
\eq
\label{eq:pressure_force_0}
 p_n(r) &=&  {1\over 6m} \ \partial^2\ {\tilde {\cal D}}_n(r) =  {1\over 6m} \ \frac{1}{r^2}\frac{d}{dr} r^2\frac{d}{dr}\ {\tilde {\cal D}}_n(r) \ , \\
~~~~~~  s_n(r) &=& -{1\over 4m} r{d\over dr} {1\over r}{d\over dr} \, {\tilde {\mathcal D}}_n(r) \ ,
\en }
and we found,
\sub{ 
\eq \label{eq:mathcal_D_r}
{\tilde {\mathcal D}}_{0}(r) &=& 2m \int{d^3\Delta\over 2E(2\pi)^3} e^{-i\vec\Delta\cdot\vec r} {\mathcal D}_{0}(t) \ , \\
{\tilde {\mathcal D}}_{2}(r) &=& 2m \int{d^3\Delta\over 2E(2\pi)^3} e^{-i\vec\Delta\cdot\vec r} {\mathcal D}_{2}(t) + \frac2m \left( \frac{d}{dr}\frac{d}{dr} -\frac2r \frac{d}{dr} \right) \int{d^3\Delta\over 2E(2\pi)^3} e^{-i\vec\Delta\cdot\vec r} {\mathcal D}_{3}(t) \ , \\
{\tilde {\mathcal D}}_{3}(r) &=& -\frac4m \left( \frac{d}{dr}\frac{d}{dr} -\frac2r \frac{d}{dr} \right) \int{d^3\Delta\over 2E(2\pi)^3} e^{-i\vec\Delta\cdot\vec r} {\mathcal D}_{3}(t) \ .
\en }
with ${\mathcal D}_{n}(t)=\sum_a {\mathcal D}_{n}^a(t)$. The detailed definitions for the form factors
$\mathcal{E}_{0,2}^a(t)$, $\mathcal{J}^a(t)$ and $\mathcal{D}_{0,2,3}^a(t)$ are shown in the Appendix \ref{sec:apdix-gffs}. The spherical components of the force (${dF_r}$, ${dF_\theta}$ and ${dF_\phi}$) acting on the infinitesimal radial area element $dS_r$  ($d\bm{S}=dS_r\bm{e}_r+dS_\theta\bm{e}_\theta+dS_\phi\bm{e}_\phi$) are given in Ref.~\cite{Panteleeva:2020ejw}. 
For completeness, we include the results in Appendix \ref {appendix:quadrupole}. As shown in Eq.(\ref{Eq:force-spherical components}) and (\ref{Eq:force-spherical components1}), in the unpolarized case, only the normal force ${dF_r}/{dS_r}$ exists inside the particle system and it only has contributions from $p_0(r)$ and $s_0(r)$. In polarized cases, both normal and tangential forces shows up, together with the higher-ordered ones, $p_2(r)$, $s_2(r)$, $p_3(r)$ and $s_3(r)$.\\

\subsection{Phenomenological light-front constituent quark}

\label{sec:model}

\begin{figure}
\resizebox{0.4\textwidth}{!}{%
\includegraphics{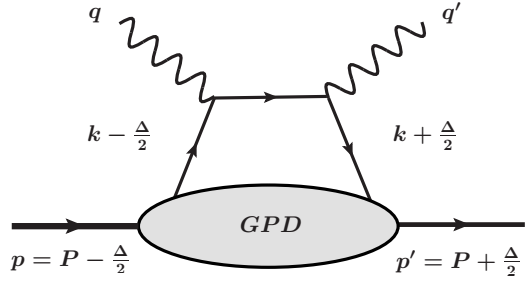}}
\caption{The s-channel handbag diagram for GPDs. The u-channel one can be obtained by $q\leftrightarrow q'$.}
\label{fig:gpd}       
\end{figure}

We know that the GFFs can be obtained from GPDs via the sum rules.
For the quark sector, one has
\sub{
\label{eq:sum-rule-q}
\eq
\label{eq:sum-rule-q-1}
\int_{-1}^1 xdx H^q_1 (x,\xi,t) &=& A^q_0(t)-\xi^2 D^q_0(t)+ {t\over 6m^2} E^q(t) + \frac13{\bar f}^q(t) \ ,   \\
\int_{-1}^1 xdx H^q_2 (x,\xi,t) &=& 2J^q(t) \ ,    \label{eq:spin-1-ji-sum-rule-q}  \\
\int_{-1}^1 xdx H^q_3 (x,\xi,t) &=& -\frac12 \left[ A^q_1(t) +\xi^2 D^q_1 (t) \right] \ ,   \\
\int_{-1}^1 xdx H^q_4 (x,\xi,t) &=& -2 \xi E^q(t) \ ,   \\
\int_{-1}^1 xdx H^q_5 (x,\xi,t) &=& {t\over 2m^2} E^q(t) + {\bar f}^q(t) \ ,
\en }
and and they are similar to the gluon ones.~\cite{Cosyn:2019aio,Polyakov:2019lbq}.
Fig. \ref{fig:gpd} illustrates the process we are
considering for GPDs in our phenomenological model. The notations are~\cite{Sun:2017gtz}
\eq
t &=&\Delta^2=(p'-p)^2=(q-q')^2 \ , \; \; Q^2=-q^2 \ ,  \nonumber \\
\xi &=&-\frac{\Delta\cdot n}{2P\cdot n}= -\frac{\Delta^+}{2P^+} \ , \ \ \abs{\xi}=\frac{\Delta^+}{2P^+}
\ , \; \; (\,|\xi\,|\le1)  \\
x &=&\frac{k\cdot n}{P\cdot n}=\frac{k^+}{P^+} \ , \ \ \ \ \ \ (-1\le x\le1) \nonumber \ ,
\en
where $n$ is a light-like 4-vector. Here $q$ is the virtual photon momentum, and $q'$ is treated as a
real one.\\

In a numerical calculation, we employ the phenomenological light-front quark model to describe the interaction
between the spin-1 $\rho$ meson and its two constitutes $u$ and ${d}$~\cite{Sun:2017gtz,Choi:2004ww}. It is based on
an effective interaction Lagrangian for the $\rho\rightarrow\bar{q}q$ vertex,
\begin{eqnarray}\label{key}
\lefteqn{
\mathcal{L}_I = -{i m_q\over f_\rho} \bar{q} \Gamma^\mu \mathbf{ \tau} q \cdot\mathbf{\rho}_\mu}
\\ \nonumber
&&= -{i \sqrt{2} m_q\over f_\rho}
\left[ \frac{\bar{u}  \Gamma^\mu u
- \bar{d}  \Gamma^\mu d }{\sqrt{2}} \rho^0_\mu
+ \bar{u}  \Gamma^\mu d  \rho^+_\mu
+ \bar{d}  \Gamma^\mu u  \rho^-_\mu
\right] ,
\end{eqnarray}
where $\rho_\mu$ is the $\rho$ meson field, $f_\rho$ is the $\rho$ decay constant,
and $\Gamma^\mu$ is a Bethe-Salpeter amplitude (BSA) describing the interaction between the meson and the
quark-antiquark pair,
\begin{eqnarray}
\Gamma^\mu=N\frac{
\gamma^{\mu} - {(k_{q}+k_{\bar q})^{\mu}}/{(M_{i,f}+2m_q)}
 }{ [ k_{q}^2-m^2_R+ \imath \epsilon] [ k_{\bar q}^2-m^2_R+ \imath \epsilon] } \ ,
\end{eqnarray}
where, for the $u$ quark contribution, the struck $u$ quark momentum is $k_u=k-\Delta/2$ and the spectator
constituent momentum is $k_s=k_{\bar{d}}=k-P$. $N$ is the normalization constant, $m_q$ and $m_R$ are the masses of
the constituent quark and the regulator, respectively.
$M_{i,f}$ are the kinematic invariant masses with subscript $i$ for initial vertex and $f$ for the final vertex,
\sub{
\begin{eqnarray}
\label{eq:vertexM:v}
M_{i}^2 = \frac{\kappa^2_\perp + m_q^2}{1-x'} + \frac{\kappa^2_\perp + m_q^2}{x'} \ , \\
M_{f}^2 = \frac{\kappa'^2_\perp + m_q^2}{1-x''} + \frac{\kappa'^2_\perp + m_q^2}{x''} \ ,
\end{eqnarray} }
with the light-front momentum fractions $x'= - k_s^+/p^+ =(1-x)/(1-\abs{\xi})$, $x''=x' p^+/p'^+ = (1-x)/(1+\abs{\xi}) $,
and
\sub{
\begin{eqnarray}
\kappa_\perp &=& k_{s\perp} - \frac{k_s^+}{p^+} p_{i\perp} \, =
(k-P)_\perp- \frac{x'}{2} {\Delta}_{\perp} \ ,  \\
\kappa'_\perp &=&
(k-P)_\perp+ \frac{x''}{2} {\Delta}_{\perp} \ . 
\end{eqnarray} }
In the ERBL regime (i.e. nonvalence regime), the relation of $-\abs{\xi}<x<\abs{\xi}$ leads to $x' > 1$, and the initial vertex becomes the non-wave-function vertex which means $M^2_i$ can get negative values.
To keep the mass square positive, we follow Ref.~\cite{Choi:2004ww} by directly
replacing $1-x'$ with $x'-1$ in Eq.~(\ref{eq:vertexM:v}) and gets
\begin{eqnarray}
\label{eq:vertexM:nv}
{\tilde M}_{i}^2 = \frac{\kappa^2_\perp + m_q^2}{x'-1} + \frac{\kappa^2_\perp + m_q^2}{x'}.
\end{eqnarray}
When both the struck and spectator constituents are on mass shells, one gets $M_{i}^2=M_{f}^2=m^2$ but ${\tilde M}_{i}^2\ne m^2$. The physics in the ERBL regime is much more complicated than that in the DGLAP one, since the creation of the $q\bar{q}$ pair involves an infinite sum of the meson contribution. The above simple method may omits the rich details. As Eq.~(\ref{eq:sum-rule-q}) shows, the GFFs $D_0(t)$ and $D_1(t)$ is bound to $\xi$ and the nonzero $\xi$ requires  the ERBL regime. It's one possible reason why our results for these two GFFs are very different with that of other models and the free theory.\\

\section{Numerical results}
\label{sec:Results}

Following our previous work on the $\rho$ meson GPDs~\cite{Sun:2017gtz,Sun:2018tmk,Sun:2018ldr}, we take the two model
parameters, the constituent mass $m_q=0.403~\gev$ and regulator mass $m_R=1.61~\gev$. The renormalization scale is about 0.5 GeV. In our LCCQM, the gluon
contributions are assumed to be absorbed into the constituent quark mass.  After summing over all the contributions
from the quark flavors, we get the total GFFs, where the 3 energy-momentum non-conserving terms are canceled,
and only 6 conserving terms are considered.\\

\begin{figure*}[t]
\centering
\subfigure[$ $]{\label{fig:gff-A0}\includegraphics[width=7.5cm]{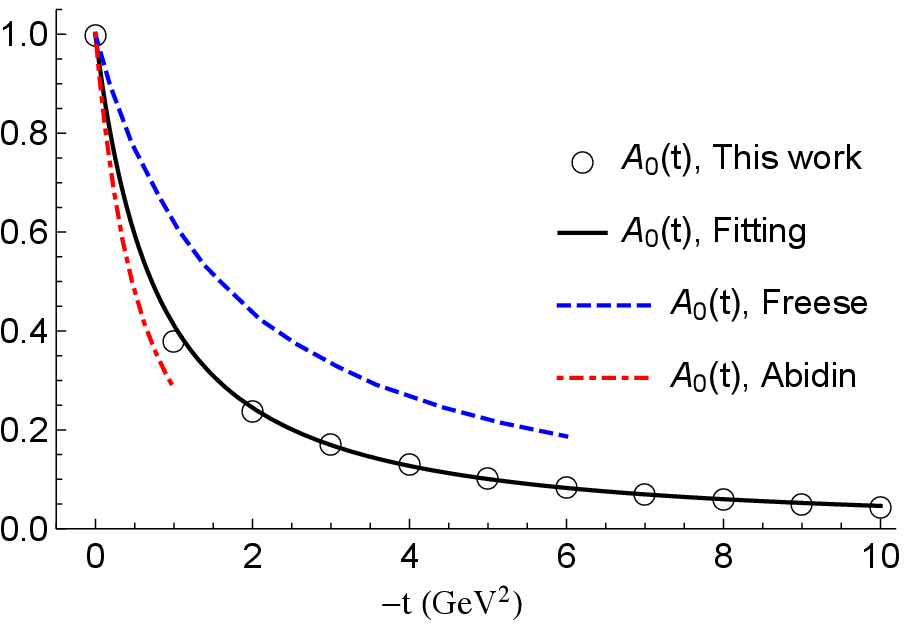}}
{\hskip 0.4cm}
\subfigure[$ $]{\label{fig:gff-A1}\includegraphics[width=7.5cm]{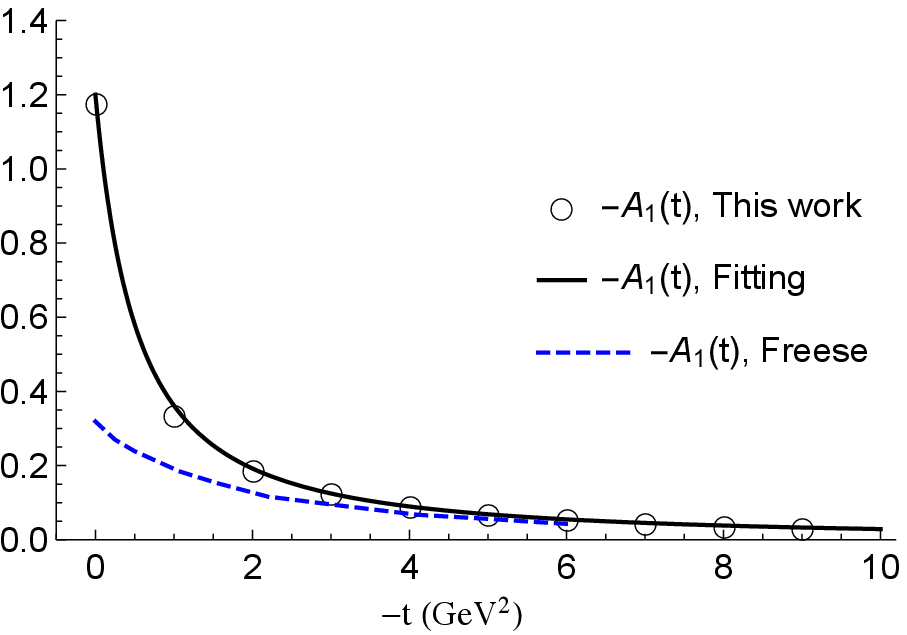}}
\subfigure[$ $]{\label{fig:gff-J}\includegraphics[width=7.5cm]{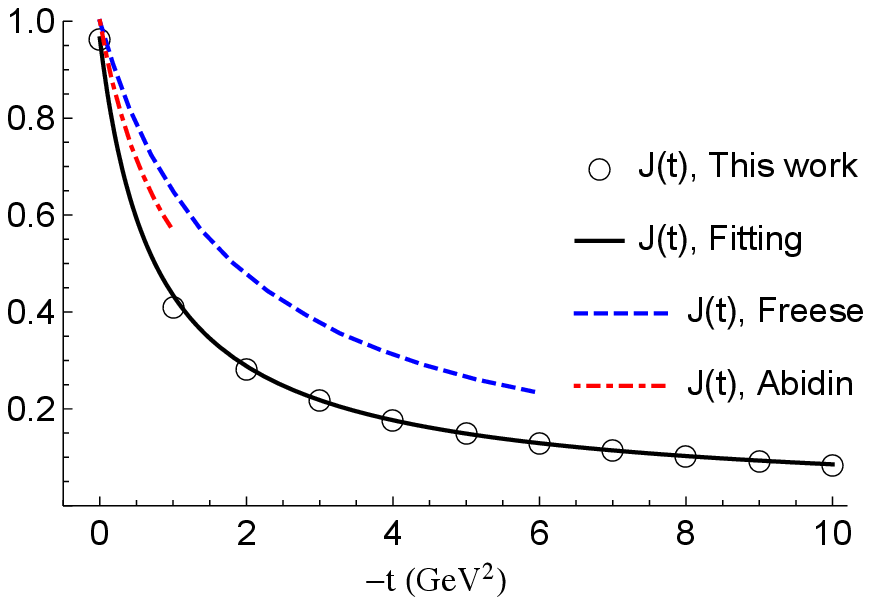}}
{\hskip 0.4cm}
\subfigure[$ $]{\label{fig:gff-E}\includegraphics[width=7.5cm]{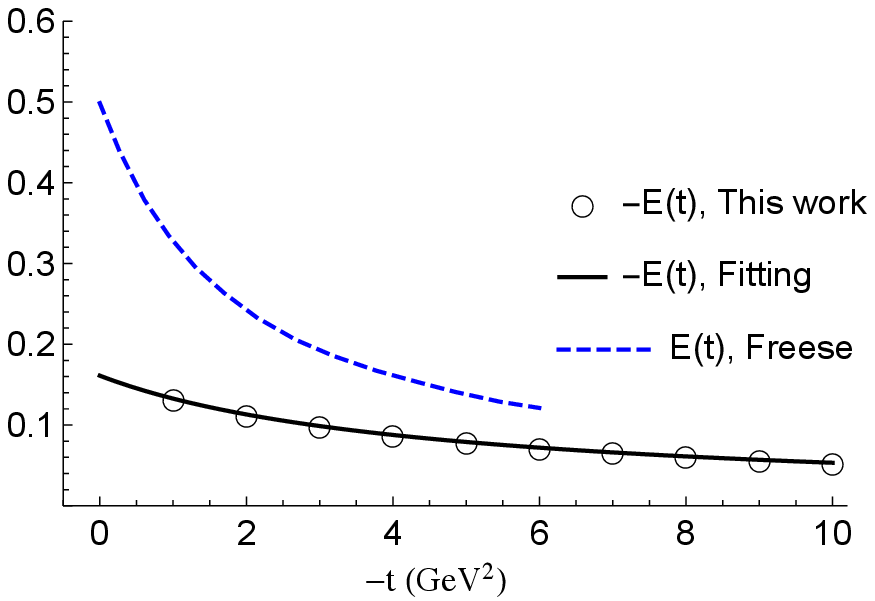}}
\subfigure[$ $]{\label{fig:gff-D0t}\includegraphics[width=7.5cm]{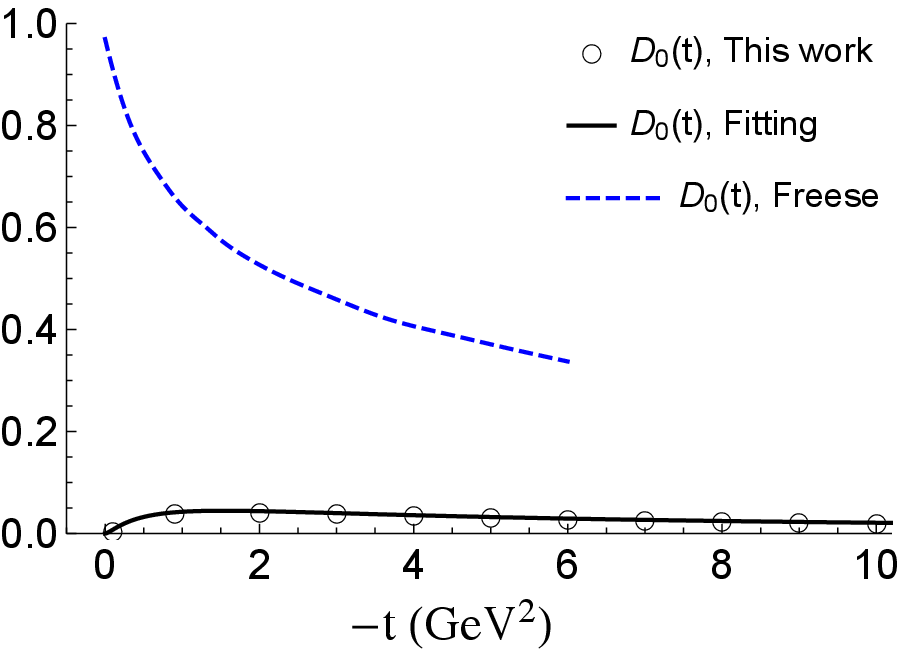}}
{\hskip 0.4cm}
\subfigure[$ $]{\label{fig:gff-D1_v4}\includegraphics[width=7.5cm]{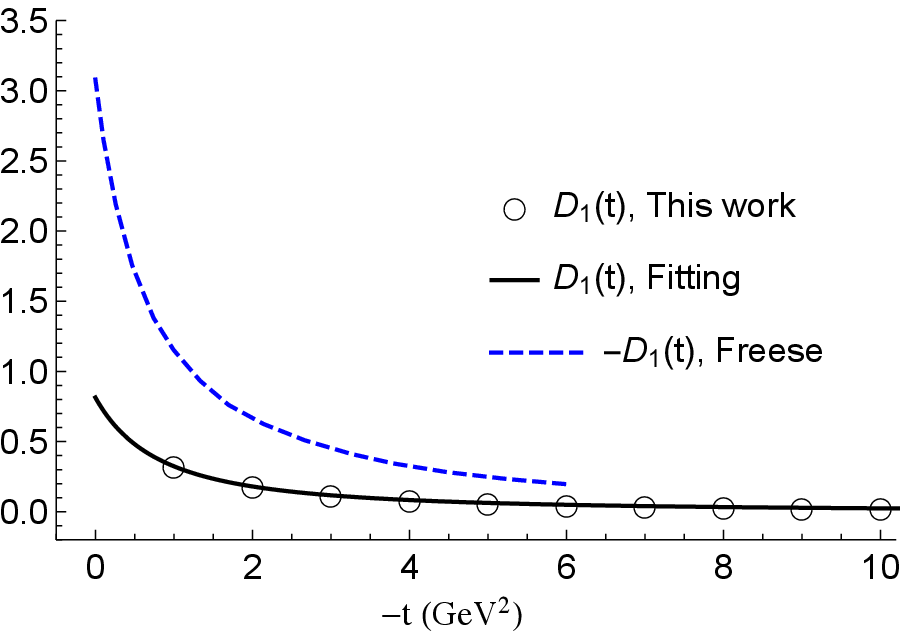}}
\caption{ Gravitational form factors (a) $A_0(t)$, $A_1(t)$, (b) $J(t)$,  $E(t)$, (c) $D_0(t)$, and (d) $D_1(t)$. The solid lines are parametric fittings and the empty circles are model results. The red dot-dashed lines are results from the AdS/QCD approach by the Abidin etc.~\cite{Abidin:2008ku} and the blue dashed lines are results from the NJL model by Freese etc.~\cite{Freese:2019bhb}. To compare the results from Ref.~\cite{Abidin:2008ku} with ours, we need take the scale $\Lambda_{\text{QCD}}=0.226$ GeV which is employed in our previous work~\cite{Sun:2017gtz}.
}
\label{fig:gff}
\end{figure*}

Our results for GFFs are shown in Fig.~\ref{fig:gff}. We know that the GFFs $A_0(t)$ and $J(t)$ are related to the
generators of the Poincare group for the mass and spin of the particle (here is the $\rho$ meson) which give the
constraints at zero-momentum transfer $A_0(0)=1$ and $J(0)=1$ \cite{Holstein:2006ud,Abidin:2008ku,Cotogno:2019xcl,Lorce:2019sbq}. Except for the cases of free particles and
Goldstone bosons,  there is no any other general principles or constraints for $D$-term~\cite{Polyakov:2018zvc}.
The results for the $D$-terms of the proton and pion have been given by different phenomenological analyses
based on the experimental data of leading order deep virtual Compton scattering (DVCS) process~\cite{Kumericki:2015lhb}.
In spin-1 case, the $D$-term is related to the GFFs $D(t)$ and $E(t)$. It is expected, with a similar approach, one
can also obtain an estimation for the mechanical properties for other spin-1 particles, in particular for deuteron,
which may be measured in the future JLab experiment~\cite{Hattawy:2019rue}.\\

In the literature, there are some other model calculations for the spin-1 GFFs before  present work.
Ref.~\cite{Abidin:2008ku} gives the results for $A_0(t)$ and $J(t)$ by applying the AdS/QCD approach, and
Ref.~\cite{Freese:2019bhb} shows the six non-zero GFFs in the NJL model. The relations
among the different notations have been explicitly discussed in Ref.~\cite{Polyakov:2019lbq}. Besides the common
constraints from the total mass ($A_0(0)=1$) and spin ($J(0)=1$), the present calculation  and the mentioned two
other approaches, however, show different decreasing $t-$dependent behaviors of the two GFFs, $A_0(t)$ and $J(t)$.
For the rest four non-zero GFFs $A_1$, $D_0$, $D_1$ and $E$, our results are also quite different from that in the NJL model~\cite{Freese:2019bhb}. One could find, in the common region of momentum transfer, the absolute value of our $A_1(t)$ ($\sim 1.2$ at $t=0$)
is much larger than the results ($\sim 0.4$ at $t=0$) in Ref.~\cite{Freese:2019bhb} and the absolute value of our $E(t)$ ($\sim 0.15$ at $t=0$)
is smaller than that ($\sim 0.5$ at $t=0$) in Ref.~\cite{Freese:2019bhb}. 
For $D_0(t)$, we get $D_0(0)\sim 0$ which is a significant difference w.r.t. the free theory $D_0(0)=1$ (without the non-minimal term)  \cite{Polyakov:2019lbq} and the chiral limit \cite{Freese:2019bhb}. One possible reason for such a difference is due to the simplification we used in the model, as discussed previously.
The other possible reason is that the quark mass is large and much away from the chiral limit.
For $D_1(t)$ we get opposite sign comparing with that from Ref.~\cite{Freese:2019bhb}.  \\

The present model result for $\rho$ meson D-term is
\eq \label{eq:D-term-spin-1_2}
D = \mathcal{D}_{0}(0)= -{D}_{0}(0)+ {4\over 3} {E}(0) = 0+{4\over 3}\cdot (-0.161)= -0.21 < 0 \ .
\en
It should be stressed that the negative value of the $D$-term satisfies the requirement for the mechanical stability
\cite{Polyakov:2018zvc}, i.e. ,
\eq
\frac{2}{3} s_0(r)+p_0(r)>0 \ .
\en
Although the t-dependent behaviors of GFFs obtained in different approaches are quite different, we find the $D$-term value
obtained from the GFFs of Ref.~\cite{Freese:2019bhb} (although it's not given explicitly in the paper) is around $-0.33$,
which is close to ours $-0.21$. \\

It is shown that the 3D Fourier transforms in the Breit frame (BF) and two-dimensional(2D) Fourier transforms in the Light-Cone (LC) frame gives different definitions of the mass radii $\langle r^2\rangle_{\text{mass}}$ \cite{Lorce:2018egm,Miller:2018ybm,Freese:2019bhb}. In the Breit frame, we have\footnote{In our previous proceeding paper \cite{Sun:2020jng}, the definition of $\langle r^2\rangle_{\text{mass}}$ in its Eq.(7) is wrong. In Ref.~\cite{Abidin:2008ku} where an AdS/QCD model calculation is preformed, its Eq.(46) defines the radius as $\langle r^2\rangle_{\text{mass}}=-6\frac{\partial A}{\partial Q^2}\Big|_{Q^2=0}$ with $Q^2=-t$.}
\eq \label{eq:radius}
\langle r^2\rangle_{\text{BF}}
&=& {\int d^3 r\, r^2 \, T^{00}({\vec r}\,)  \over \int d^3 r \, T^{00}({\vec r}\, ) }
=\frac 1m \int d^3{\bf  r} \ r^2\ T^{00}({\vec r}\,) \nonumber \\
&=& 
6 {d A_0(t) \over d t} \bigg|_{t\rightarrow0} +
\frac{1}{m^{2}}\left[-\frac{7}{4}A_{0}(0)+\frac{1}{2}A_{1}(0)+\frac{3}{2}D_{0}(0)+2J(0)-E(0)\right] \ , 
\en
Eq.(\ref{eq:radius}) is equivalent to Eq. (36) of Ref. \cite{Cosyn:2019aio}.
In the Light-Cone frame, it is obtained in Ref.~\cite{Freese:2019bhb} as
\eq
{\langle r^2\rangle_\text{LC}}
&=& 4\frac{dA_{0}\left(t\right)}{dt}\bigg|_{t=0}+\frac{1}{3m^{2}}\left[2A_{0}\left(0\right)+A_{1}\left(0\right)-2J\left(0\right)+2E\left(0\right)\right] \ .
\en
As discussed in Ref.~\cite{Miller:2018ybm} and \cite{Freese:2019bhb},  the 3D spatial distributions in the Breit frame are found not invariant under Lorentz boosts. While the relativistic corrections are intrinsically accounted for in the Light-Cone frame and therefore the physical meaning of the radius of 2D transverse distributions in Light-Cone is more clear. 
The numerical results of these radii are list in Table.~\ref{radiusQ} in comparing with other model predictions. In present work, we have $\sqrt{\langle r^2 \rangle_\text{BF}} =0.53$ fm and $\sqrt{\langle r^2\rangle_\text{LC}}=0.41$ fm. In both frames, the mass radii are smaller than the charge radius ($0.72$ fm) from our previous calculation.
This feature is reasonable and consistent with the nucleon case~\cite{Goeke:2007fp,Bezginov:2019mdi}.
Besides, we find both of the radius and the $D$-term are sensitive to the constituent quark mass $m_q$. As binding
energy is approaching zero ($m_q$ is approaching to the half of total $\rho$ meson mass), the radius increases rapidly
and the absolute value of $D$-term decreases close to zero. It means the bound system is getting looser or even falls
apart in the view of the constituent quark model.
It is consistent with the observation in Ref.~\cite{Hudson:2017oul} that the value of $D$-term vanishes in the free Dirac fermions case. \\

The gravitational quadrupole moment, 
\eq
{\mathcal Q}_{\text{mass}} = -\frac1m\left[ -A_0(0) +\frac12 A_1(0) +2J(0) -E(0) \right] = -0.0322 \; [m_\rho\text{-fm}^2], 
\en
 in present model. It is consistent with that from the NJL model prediction ($-0.0224\;[m_\rho\text{-fm}^2]$) in Ref.~\cite{Freese:2019bhb}. The quadrupole moments of mass and charge are close under the comparable units. The same (negative) sign implies the mass and charge distributions are synchronous when the particle becomes polarized.\\

\begin{table*}
\caption{\label{radiusQ} Mean squared mass radius, mass and quadrupole moment of $\rho$ meson by this work, the NJL model~\cite{Freese:2019bhb} and the AdS/QCD model~\cite{Abidin:2008ku}, respectively. All radii are in fm, the mass
quadrupole moment is in units of $m_\rho$-fm${}^2$,and the electric quadrupole moment is in $e$-fm. In Ref.~\cite{Abidin:2008ku}, it is not specified in which frame the definition of radii is given, and the mass definition differ from that the Breit frame and light cone prescriptions used in the present work and Ref.~\cite{Freese:2019bhb}. 
}
\begin{center}
	\begin{tabular}{ccccc}
	\hline
	\hline \\
	$ $ & $\sqrt{\la r^2 \ra_{\text{mass}}}$ & $\sqrt{\la r^2 \ra_{\text{elec.}}}$ & ${\mathcal Q}_{\text{mass}}$ & ${\mathcal Q}_{\text{elec.}}$ \\ \hline \\
	$\text{AdS/QCD~\cite{Abidin:2008ku}}$ & $0.46$ & $0.73$ & $ $ & $ $  \\
	$\text{NJL~\cite{Freese:2019bhb}, Briet frame}$ & $0.45$ & $0.67$ & $-0.0224$ & $-0.0200$  \\
	$\text{NJL~\cite{Freese:2019bhb}, Light Cone}$ & $0.32 $ & $0.45$ & $ $ & $ $  \\
	$\text{this work, Briet frame}$ & $0.53$ & $0.72$ & $-0.0322$ & $-0.0212$  \\
	$\text{this work, Light Cone}$ & $0.41$ & $ $ & $ $ & $ $  \\
	\hline
	\hline
	\end{tabular}
\end{center}
\end{table*}

In principle, one can calculate the static EMT $T^{\mu\nu}(r)$ (also the energy density and pressure)
straightforwardly from the obtained GFFs with Fourier transformation. However, the integrals may not converge if
the GFFs drops slowly w.r.t. the momentum transfer square $t$. According to the analyses of pQCD and AdS/CFT, at
the large momentum transfer ($-t=Q^2\rightarrow \infty$), the six GFFs decease roughly with the following power
respectively~\cite{Abidin:2008ku},
\eq
\label{eq:GFFsLargeQ}
(A_0 \ , \ D_0 \ , \ J \ , \ E) \sim 1/{t}^2 \ , \quad (A_1 \ , D_1)   \sim 1/{t}^3 \ .
\en
In the nucleon case, Ref.~\cite{Burkert:2018bqq} adopts an assumption that its GFFs behave $d_1(t) \sim t^{-3}$,
and the converged results are obtained.\\

Because of the limited capability of our LCCQM, especially at the large momentum transfer region,  we believe that a
modification of our model results in the large momentum transfer region is needed. To simulate the $t$-dependent
behaviors of the obtained GFFs, we consider the forms like
\eq \label{eq:fit0}
a \left( 1- {t\over b}\right)^c
\en
to present our numerical results at a momentum transfer region, $0<-t<10~\gev^2$, and we find five of the six GFFs are
approximately described by
\sub{
\label{eq:fit1}
\eq
A_0(t) &=& { \left(1-0.996 \, t \right)^{-1.28}} \ , \\
A_1(t) &=& {-1.20 \left(1-{t\over 0.73} \right)^{-1.38}} \ , \\
D_1(t) &=& {0.814  \left(1-{t\over 1.32}\right)^{-1.64}} \ , \\
J(t) &=& {0.965  \left(1-{t\over 0.68}\right)^{-0.877}} \ , \\
E(t) &=& {-0.161  \left(1- {t\over 4.2}\right)^{-0.909}} \ .
\en }
Due to the limit $\abs{\xi}\leqslant 1/\sqrt{1-4M^2/t}$, there are some small oscillation in the numerical
result of the GFF $D_0(t)$ at $0<-t<3~\gev^2$. After a carefully check of its $t-$dependent behavior, we find it
is oscillating around a curve that can be simulated by
\eq \label{eq:fit2}
\left( 1- {t\over b}\right)^c \left(1-{d \over t} \right)^e \ .
\en
Thus, for $D_0(t)$, we get
\eq \label{eq:fit3}
D_0(t) = { \left(1-{0.97\over t}\right)^{-1.97}} \left(1-{t\over 0.14}\right)^{-0.86}\ .
\en

It turns out, unlike the nucleon case in Ref.~\cite{Burkert:2018bqq},  neither our GFFs results in Eqs. (\ref{eq:fit1})
and (\ref{eq:fit3}) nor the pQCD predictions in Eq. (\ref{eq:GFFsLargeQ}) drop fast enough to give converging results for
$T^{\mu\nu}(r)$ (see Eq.~(\ref{eq:static_EMT})). This issue has
already been pointed out and discussed for the cases of pion meson~\cite{Hudson:2017xug} and nucleon \cite{Lorce:2018egm}.  One possible
reason is that the integrals that defining $T^{\mu\nu}(r)$ is subjected to the relativistic corrections. For the spin-0
case, Ref.~\cite{Hudson:2017xug} estimated the relativistic corrections by a way of smearing out the point-particles
(a delta function is replaced by a Gaussian function). It is believed that the relativistic corrections
($\delta_{\text {rel}} \equiv 1/(2m^2R_h^2)$) is negligible when $mR_h \gg 1$ where $m$ is the hadron mass and $R_h$ is
the hadron size. For the light meson pion, its $\delta_{\text {rel}}=220\%$ ~\cite{Hudson:2017xug}. Although the
$\rho$ meson is spin-1 hadron, we may simply "borrow" the argument for pion to roughly estimate how large the
relativistic corrections are for the case of the $\rho$ meson. With our model estimation of the radius
($\sqrt{\la r \ra^2_{\text{grav}}} \sim 0.53$~fm), one gets $\delta_{\text {rel}} \sim 12\%$ which is not important and we believe
that the concept of the 3D densities is applicable for the $\rho$ meson. So far, there is no experimental data for
the $\rho$ meson radius, the future experimental information about its size would be essential for our estimate.\\

To proceed with a modification of our phenomenological model calculation, particularly in the large $t$ region,
we introduce a Gaussian form wave package, as we did in previous work~\cite{Sun:2018tmk}, to suppress the contribution
from the high energy region. It's reasonable since only limited values of $t$ can be measured in the experiments. Choosing
the Gaussian form wave package originates from the observation that a hadron is an extended object and is smeared out in
space~\cite{Diehl:2002he}. It should be stressed that the value of $D$-term is not affected by this modification and
by the consideration of the relativistic corrections since it's defined by the value of GFFs at the zero momentum
transfer~\cite{Hudson:2017xug,Polyakov:2018zvc}.\\

In our previous study for the $\rho$-meson transverse distributions in the 2D impact parameter
space~\cite{Sun:2018tmk}, we introduced a 2D Gaussian form wave packet in both of the incoming and outgoing states
in order to avoid the similar divergences, and we found the appropriate value for the wave packet width being around
$\sigma = 1 \sim 2 \gev^{-1}$. This treatment was also pointed out and employed in other calculations of nucleon GPDs
in the impact parameter space~\cite{Diehl:2002he}. Here, a similar 3D Gaussian form wave packet,
\eq
e^{-{{\vec \Delta}^2}{\sigma_0^2}/4} \ ,
\en
is adopted (with width $\sigma_0 \sim 2 ~\gev^{-1}= 0.39$~fm) to carry out the calculation for the spatial
distributions. As a result, the expressions for the energy densities are modified to be
\sub{
\label{eq:energy-densities-Gauss}
\eq
\varepsilon^{(\sigma_0)}_0(r) &=&  2m^2 \int{d^3\Delta\over 2E(2\pi)^3}
e^{-i\vec\Delta\cdot\vec r -{\vec\Delta^2}{\sigma_0^2}/4}{\mathcal E}_{0}(t) \ , \\
\varepsilon^{(\sigma_0)}_2(r) &=& -{r} {d\over dr} {1\over r} {d\over dr} \int{d^3\Delta\over 2E(2\pi)^3}
e^{-i\vec\Delta\cdot\vec r -{\vec\Delta^2}{\sigma_0^2}/4} {\mathcal E}_{2}(t) \ .
\en }
Then, the spin distribution is modified to be
\eq \label{Eq:static-EMT-T0k-3}
J_a^{i(\sigma_0)}({r},\sigma^\prime,\sigma)
&=& J_a^{i(\sigma_0)}(\vec{r},\sigma^\prime,\sigma)  \nonumber \\
&=&{\hat S}^j_{\sigma^\prime\sigma}  \int {d^3 \Delta \over  (2\pi)^3 }
e^{-i \vec \Delta \cdot \vec r -{\vec \Delta^2}{\sigma_0^2}/4}  \bigg[ \Big(  \mathcal{\hat J}^a(t)
+\frac23 t {d \mathcal{\hat  J}^a(t) \over dt}  \Big) \delta^{ij}
\nonu
+ \Big( \Delta^i\Delta^j - \frac13 \vec \Delta^2 \delta^{ij}  \Big) {d \mathcal{\hat J}^a(t) \over dt}  \bigg]\;,
\en
where $\mathcal{\hat J}^a(t)=\frac{m}{E} \mathcal{J}^a(t)$ and the first step in the above equation is based on
the observation that $J_a^i$ depends on $\vec{r}$ only through $r=|\vec{r}|$. Moreover, the function
${\tilde {\mathcal D}}_{n}(r)$ in Eq.~(\ref{eq:mathcal_D_r}), which defines the distributions of pressure and shear
force, is modified as
\sub{
\eq \label{eq:ft2D}
{\tilde {\mathcal D}}^{(\sigma_0)}_{0}(r) &=& 2m \int{d^3\Delta\over 2E(2\pi)^3} e^{-i\vec\Delta\cdot\vec r -{\vec\Delta^2}{\sigma_0^2}/4} {\mathcal D}_{0}(t) \ , \\
{\tilde {\mathcal D}}^{(\sigma_0)}_{2}(r) &=& 2m \int{d^3\Delta\over 2E(2\pi)^3} e^{-i\vec\Delta\cdot\vec r -{\vec\Delta^2}{\sigma_0^2}/4} {\mathcal D}_{2}(t) + \frac2m \left( \frac{d}{dr}\frac{d}{dr} -\frac2r \frac{d}{dr} \right) \int{d^3\Delta\over 2E(2\pi)^3} e^{-i\vec\Delta\cdot\vec r-{\vec\Delta^2}{\sigma_0^2}/4} {\mathcal D}_{3}(t) \ , \\
{\tilde {\mathcal D}}^{(\sigma_0)}_{3}(r) &=& -\frac4m \left( \frac{d}{dr}\frac{d}{dr} -\frac2r \frac{d}{dr} \right) \int{d^3\Delta\over 2E(2\pi)^3} e^{-i\vec\Delta\cdot\vec r -{\vec\Delta^2}{\sigma_0^2}/4} {\mathcal D}_{3}(t) \ .
\en }
Correspondingly, $p_n(r)\rightarrow p_n^{(\sigma_0)}(r)$ and $s_n(r)\rightarrow s_n^{(\sigma_0)}(r)$.\\

After summing over all the partons (only quarks here) in
Eq. (\ref{Eq:static-EMT-T0k-3}), one has
\eq
J^{i(\sigma_0)}_{\sigma^\prime\sigma}(\vec{r}\,)= \sum_a J_a^{i(\sigma_0)}(\vec{r},\sigma^\prime,\sigma) \ ,
\ i,\sigma^\prime,\sigma=x,y,z \ .
\en
In the Cartesian basis, average over all polarizations and spatial directions, one further has
\eq
J^{(\sigma_0)}(r)\equiv \frac{1}{\text{Tr}[\hat{\bf S}^2]} \sum_{\sigma'\sigma i} {\hat S}^i_{\sigma^\prime\sigma}
J^{i(\sigma_0)}_{\sigma^\prime\sigma}(\vec{r}\,) = i { J}^{x(\sigma_0)}_{yz}(r)\ ,
\en
which is a real quantity and is the 3D spatial spin distribution of the $\rho$ meson. \\

\begin{figure}[t]
\centering
\subfigure[$ $]{\label{fig:T00_e0e2}\includegraphics[width=7.5cm]{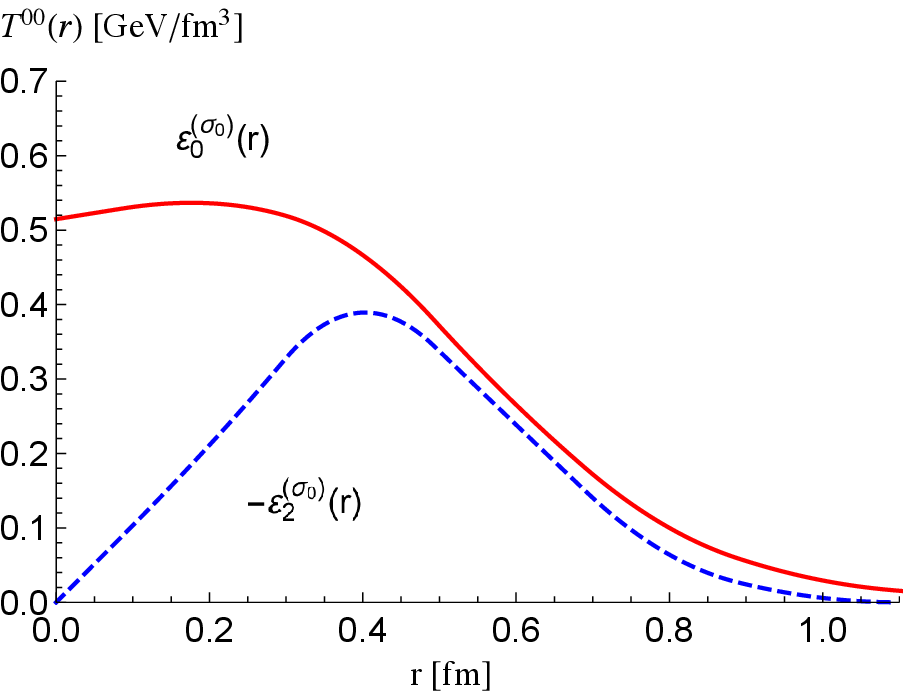}}
{\hskip 0.5cm}
\subfigure[$ $]{\label{fig:T00_e0e2_r2}\includegraphics[width=7.5cm]{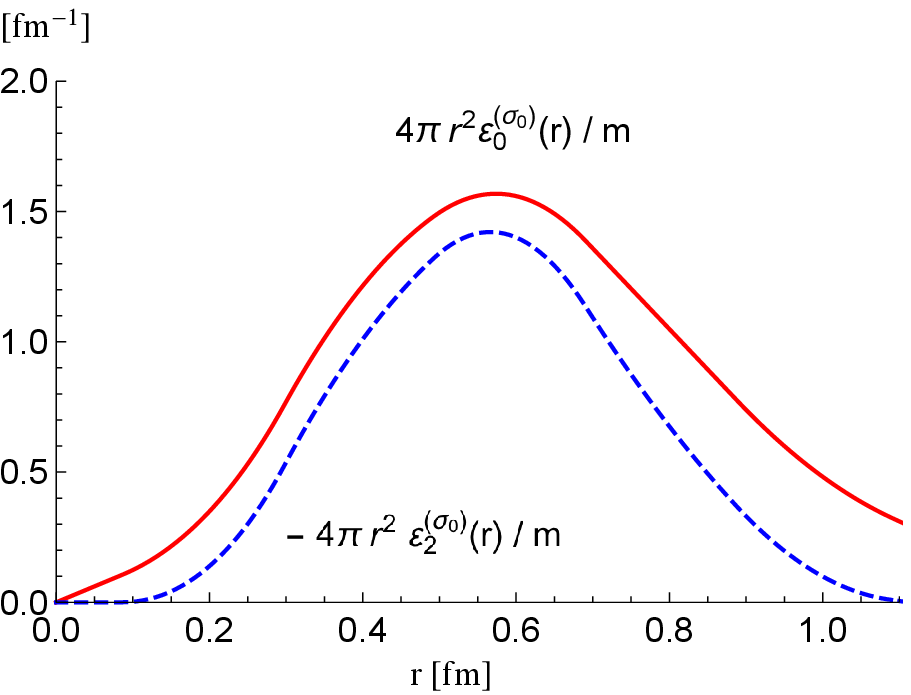}}
\caption{ Energy densities (a) ${\cal \varepsilon}^{(\sigma_0)}_0(r)$,
(b) ${\cal \varepsilon}^{(\sigma_0)}_2(r)$ with $\sigma_0 = 2 ~\gev^{-1}$. } \label{fig:T00r}
\end{figure}

With those preparations, the energy densities calculated from the GFFs in Eqs. (\ref{eq:fit1}) and (\ref{eq:fit3})
are shown in Fig. \ref{fig:T00r}. In Fig. \ref{fig:T00_e0e2_r2}, the normalization is changed to the form of
$4\pi r^2 {\cal \varepsilon}^{(\sigma_0)}_0(r) /m$, which gives 1 after averaging over the polarizations and
integrating over the whole radial space. The higher-order term ${\cal \varepsilon}^{(\sigma_0)}_2(r)$ doesn't
contribute to the energy distributions in the unpolarized case. Its
negative value indicates that the mass or energy distribution would deviate from the center because of the polarization
effect. 
As shown in Table \ref{radiusQ}, the values of charge and mass quadrupole moments in our work and Ref.\cite{Freese:2019bhb} are all negative. In the classical picture, a negative quadrupole moment corresponding to an oblate ellipsoid distribution. Thus the charge and mass distributions are consistent in shape. \\

\begin{figure*}[t]
\centering
\includegraphics[width=8cm]{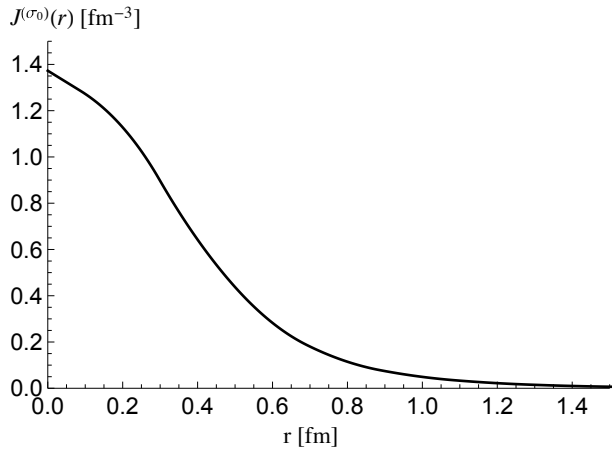}
\caption{ Spin densities $J^{(\sigma_0)}(r)$ with $\sigma_0 = 2 ~\gev^{-1}$.  } \label{fig:Jr}
\end{figure*}

The result for the spin distribution is shown in Fig.~\ref{fig:Jr}.
In our previous work with LCCQM, we obtain the fraction of spin carried by the constituent quark and antiquark in
$\rho$ meson is $86\%$~\cite{Sun:2018ldr}. The rest part is believed to come from the orbital angular momentum (no
gluon in our model)~\cite{Hoodbhoy:1998yb,Lorce:2011kd,Hatta:2011ku,Kanazawa:2014nha}.  \\

\begin{figure}[t]
\centering
\subfigure[$ $]{\label{fig:Tij_p0s0}\includegraphics[width=7.5cm]{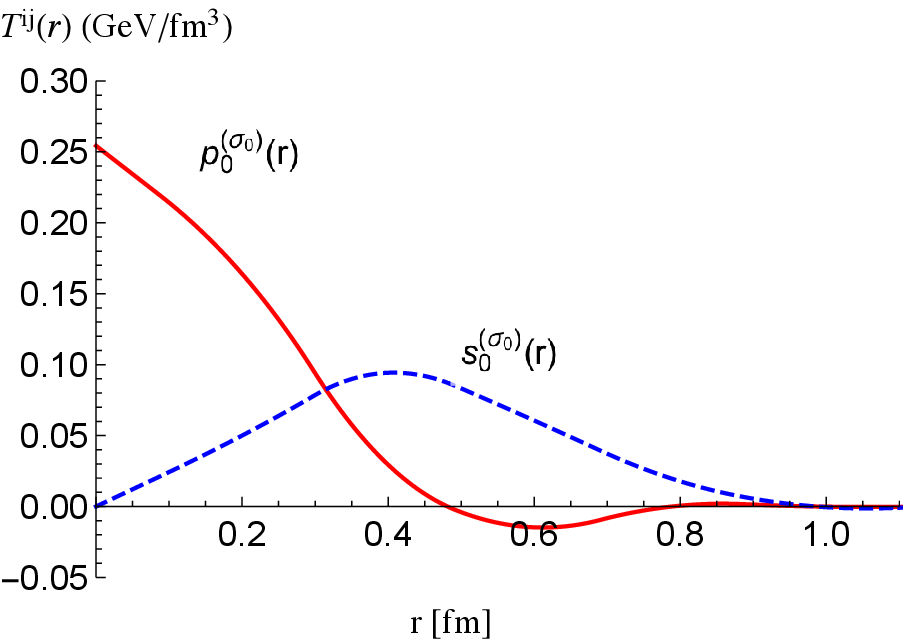}}
{\hskip 0.5cm}
\subfigure[$ $]{\label{fig:r2p0rSigma2}\includegraphics[width=7.5cm]{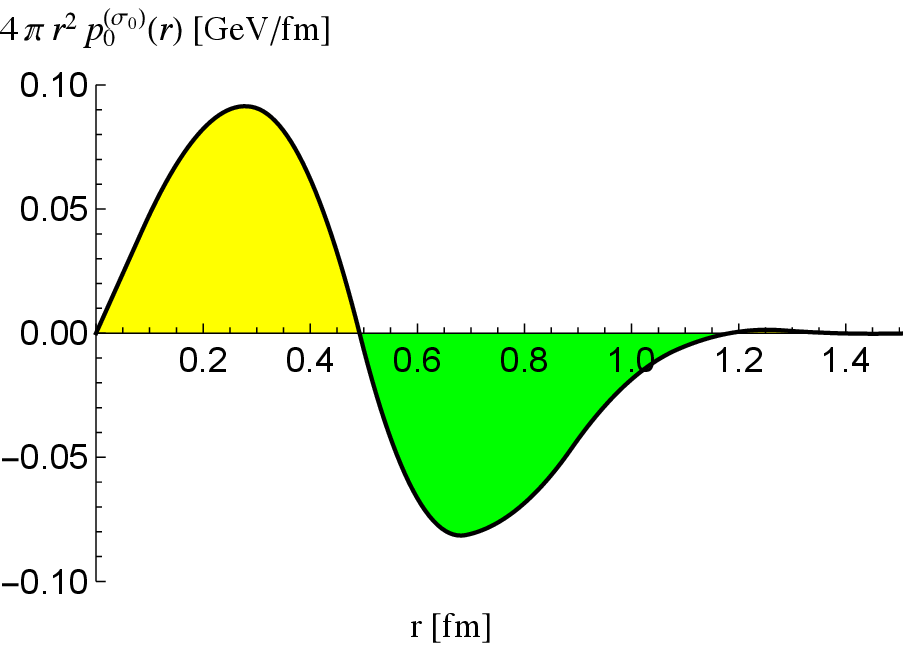}}
\subfigure[$ $]{\label{fig:Tij_p2s2}\includegraphics[width=7.5cm]{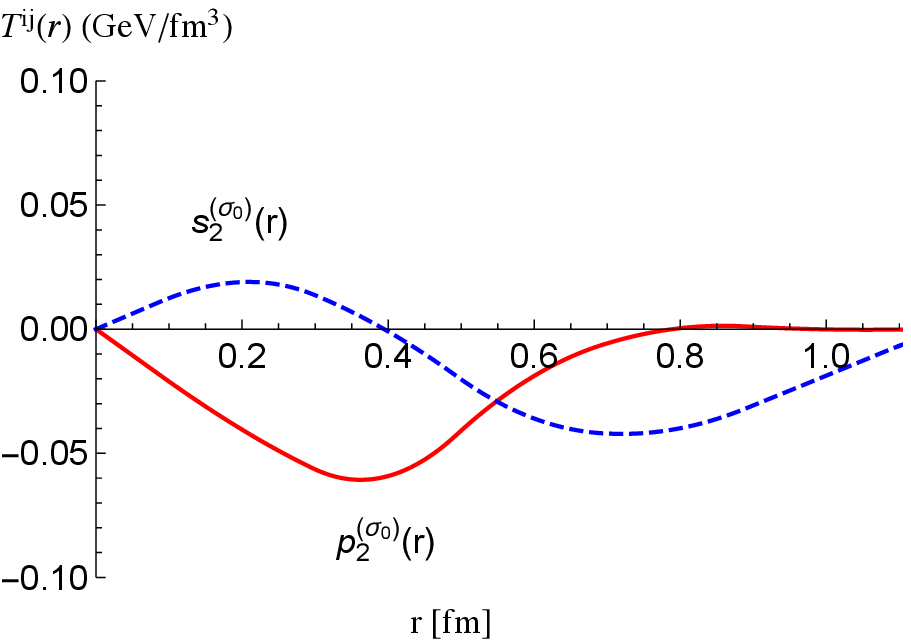}}
{\hskip 0.5cm}
\subfigure[$ $]{\label{fig:Tij_p3s3}\includegraphics[width=7.5cm]{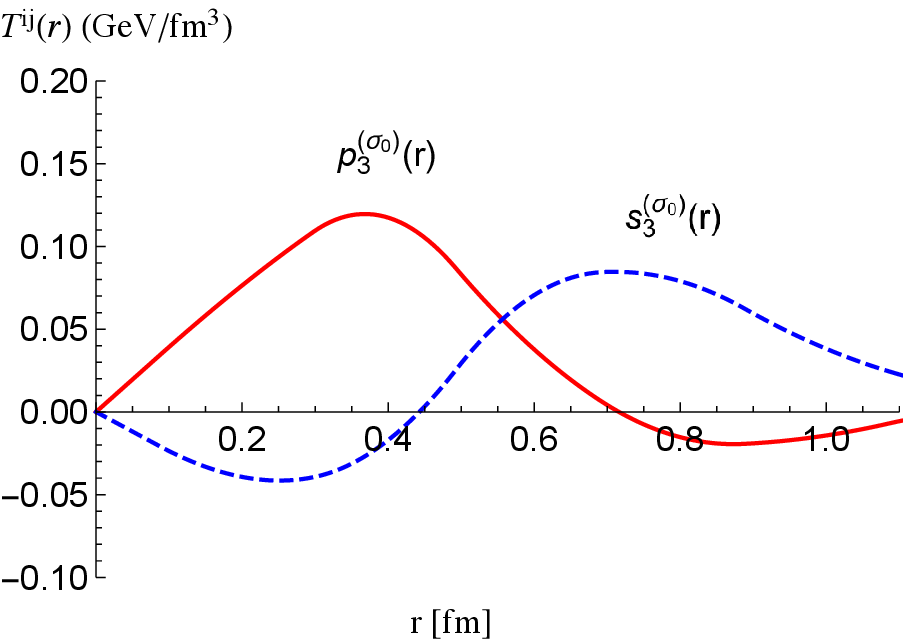}}
\caption{ The pressure and shear forces functions with $\sigma_0 = 2 ~\gev^{-1}$. } \label{fig:Tijr}
\end{figure}

Finally, our results for pressures  $p^{(\sigma_0)}_n$ and shear forces $s^{(\sigma_0)}_n$ are shown in Fig. \ref{fig:Tijr}.
To the best of our knowledge, there are some model calculations for spin-0 and spin-1/2 nucleon, but none for the
spin-1 particle before. In the unpolarized case, only the first order pressure and shear force contribute to $T^{ij}$
in Eq.~(\ref{eq:thetauvij_2}) and the normal force $dF_r$ in Eq.~(\ref{Eq:force-spherical components}). So far in almost all model studies for different spin cases, it is found that the
(unpolarized) pressure is positive in the inner region and negative in the outer region. Under the present convention,
we know that the positive sign means repulsion towards outside and the negative sign means attraction towards inside.
However, the specific relationship between these values and the strong force remains obscure. As one can see,
$p^{(\sigma_0)}_0(r)$ changes its sign for the first time at around $r\sim 0.5$~fm, which is roughly the
gravitational radius. Physically, the sign-changing means the forces change from "stretching"  to "squeezing".
This phenomenon is the same as in the nucleon case~\cite{Polyakov:2018zvc}.
In the polarized case, the higher-ordered pressures $p_2(r)$, $p_3(r)$ and shear forces $s_2(r)$, $s_3(r)$, contributes to both normal and tangential forces, as shown in Eq.~(\ref{Eq:force-spherical components}) and (\ref{Eq:force-spherical components1}). The size of tangential forces are proportional to $p_2(r)+\frac23 s_2(r)$ which keeps negative as shown in Figure~\ref{fig:Tij_p2s2}. The final sign of $dF_\theta$ and $dF_\phi$ is, however, also dependent on the spherical quadrupole tensor elements as shown in Appendix \ref {appendix:quadrupole}.  In the large$-N_c$ limit with the baryon as chiral soliton, it is found that $p_2(r)=s_2(r)=0$ for the $\Delta$ baryon $(J=3/2)$~\cite{Panteleeva:2020ejw}, which is a very interesting prediction.
At the region, $r \ge 1$ fm, the
pressures and shear forces are all quickly approaching to zero with small oscillations w.r.t. the center region
values. There always exist oscillations in this approach and the oscillations depend on what the value of $\sigma_0$
is used in our numerical calculation (larger $\sigma_0$ generates stronger suppression on the amplitudes of the
oscillations). There is still no constraint on how many times the changing would happen and no explanation for the
meaning of those numbers. It is of great interest to have further study and to answer these intriguing
questions. Nevertheless, here we present the first model estimation for the pressures and shear forces for the
spin-1 $\rho$ meson. Similar distributions may indicate some common properties of the strong force in forming
the hadron systems.

\section{Summary}
\label{sec:Summary}

In this work, we extend our previous approach on the $\rho$ meson GPDs with the phenomenological light-front
constituent quark model to its GFFs and further to the distributions of pressure and shear forces. For the GFFs
$A_0$ and $J$ which are related to the mass and spin, our model estimations are consistent with the result of other
approaches, such as the NJL model and AdS/CFT {\it etc.}. For the rest four GFFs, there are no specific constraints
as the mass and spin cases, and the results from different approaches have large discrepancies even with opposite
signs. Moreover, the $D$-term is given through our calculated GFFs, and it is estimated to be $-0.21$. The negative
value satisfies the stability condition. We also calculate the distributions of energy, pressure, spin, and shear forces. The results for the mass radius and quadrupole moment also agree with previous calculations in the NJL model and the AdS/QCD model {\it etc.}
Since the LCCQM works well mainly within the low momentum transfer regions, we consider a Gaussian wave package during the
Fourier transforms to suppress the contributions from large momentum transfer regions. Thus, the present results should be
considered as a qualitative estimation. We expect that our results may provide some hints for the understanding of
the mechanical properties, especially in the case of spin-1 hadron.\\

\section{Acknowledgement}\label{Sec:sec5}\par\noindent\par

We acknowledge helpful conversations with Maxim V. Polyakov and Julia Yu. Panteleeva. This work is supported by the National Natural Sciences
Foundations of China under the grant Nos. 11975245, 11521505, 11565007, 11635009, and 11947228, the Sino-German CRC 110
by NSFC under Grant No.11621131001, the Key Research Program of CAS, Grant No. Y7292610K1, the IHEP Innovation Fund
No. Y4545190Y2 and China Postdoctoral Science Foundation under Grant No. 2019M662316.

\appendix

\section{Definition for the form factors}
\label{sec:apdix-gffs}
The definitions for the form factors used in this work are
\label{eq:gffs}
\eq
\label{eq:E0}
\mathcal{E}^a_{0}(t)
&=&
 A^a_0(t) + {1\over4} {\bar f}^a(t) - {1\over2} {\bar c}^a_0(t)
\nonumber \\
&&
+{t\over {12}m^2} \Bigl[ -{5} A^a_0(t)  +3 D^a_0(t)+ 4J^a(t) -2E^a(t) +A^a_1(t) +{1\over2} {\bar f}^a(t)
+ {\bar c}^a_0(t)+ \frac12 {\bar c}^a_1(t) \Bigr]
\nonumber \\
&&
-{t^2\over {24}m^4} \Bigl[ -A^a_0(t)  + D^a_0(t) + {2}J^a(t) - 2E^a(t) + A^a_1(t) +  {1\over2}D^a_1(t)
+ \frac14 {\bar c}^a_1(t) \Bigr]
+{t^3\over {192}m^6}  \Bigl[ A^a_1(t) +  D^a_1(t)  \Bigl] , \ \\ [0.7em]
\label{eq:E2}
\mathcal{E}^a_{2}(t)
&=&
-A^a_0(t) + {2}J^a(t)  -E^a(t)  + \frac12A^a_1(t) + \frac14 {\bar f}^a(t)+\frac12 {\bar c}^a_0(t)+ \frac14 {\bar c}^a_1(t)
\nonumber \\
&&
- {t\over 4m^2} \Bigl[ -A^a_0(t)  + D^a_0(t) + 2J^a(t) - 2E^a(t) + A^a_1(t) +  {1\over2}D^a_1(t) + \frac{1}{4} {\bar c}^a_1(t) \Bigr]
\nonumber \\
&& + {t^2\over 32m^4}  \Bigl[ A^a_1(t) +  D^a_1(t)  \Bigl]  \ , \\ [0.5em]
\label{eq:J}
\mathcal{J}^a(t)&=&
 J^a(t) + {1\over2} {\bar f}^a(t) -{t\over4m^2} \Bigl( J^a(t)   -  E^a(t) \Bigl) \ . \\
\label{eq:D0}
\mathcal{D}^a_0(t)&=&
-D^a_0(t)+{4\over3} E^a(t) +{t\over12m^2} \Bigl[ 2D^a_0(t)-2 E^a(t)+ D^a_1(t) \Bigl] -  {t^2\over48m^4}  D^a_1(t) \ , \\
\label{eq:D2}
{\mathcal D}^a_2(t)&=& -E^a(t) \ , \\
\label{eq:D3}
{\mathcal D}^a_3(t)&=&  \frac14 \Bigl[ 2D^a_0(t)-2 E^a(t)+ D^a_1(t) \Bigl] - {t \over 16m^2}   D^a_1(t) \ .
\en
When sum over all partons, the momentum-energy non-conserving terms, ${\bar f}^a$ and ${\bar c}_{0,1}^a$, will drop and
they have no contribution.

\section{Quadrupole tensor elements}\label{appendix:quadrupole} 

The spherical components of the force acting on the infinitesimal radial area element $dS_r$  ($d\bm{S}=dS_r\bm{e}_r+dS_\theta\bm{e}_\theta+dS_\phi\bm{e}_\phi$) read~\cite{Panteleeva:2020ejw}:
\eq
\label{Eq:force-spherical components}
	\frac{dF_r}{dS_r}&=&p_0(r) + \frac 23 s_0(r)+\hat Q^{r r}  \left(p_2(r)+ \frac 23 s_2(r)+p_3(r)+\frac 23 s_3(r)\right) , \\	
	\label{Eq:force-spherical components1} 
	\frac{dF_\theta}{dS_r}&=& \hat Q^{\theta r}  \left(p_2(r)+ \frac 23 s_2(r)\right), \quad \frac{dF_\phi}{dS_r}= \hat Q^{\phi r}  \left(p_2(r)+ \frac 23 s_2(r)\right).
\en 
Some of the spherical quadrupole tensor elements involved in Eq.~(\ref{Eq:force-spherical components}) and (\ref{Eq:force-spherical components1}) are,
\eq
\hat{Q}^{rr} & = & \left(\hat{Q}^{xx}\mathrm{cos}^{2}\phi+\hat{Q}^{xy}\mathrm{sin}2\phi+\hat{Q}^{yy}\mathrm{sin}^{2}\phi\right)\mathrm{sin}^{2}\theta+\hat{Q}^{xz}\mathrm{sin}2\theta\mathrm{cos}\phi+\hat{Q}^{yz}\mathrm{sin}2\theta\mathrm{sin}\phi+\hat{Q}^{zz}\mathrm{cos}^{2}\phi\;,\\
\hat{Q}^{\theta r} & = & \left(\hat{Q}^{xx}\mathrm{cos}^{2}\phi+\hat{Q}^{xy}\mathrm{sin}2\phi+\hat{Q}^{yy}\mathrm{sin}^{2}\phi -\hat{Q}^{zz} \right)\mathrm{sin}\theta\mathrm{cos}\theta+\left(\hat{Q}^{xz}\mathrm{cos}\phi+\hat{Q}^{yz}\mathrm{sin}\phi\right)\mathrm{cos}2\theta\;,\\
\hat{Q}^{\phi r} & = & \left(\left(\hat{Q}^{yy}-\hat{Q}^{xx}\right)\mathrm{sin}\phi\mathrm{cos}\phi+\hat{Q}^{xy}\mathrm{cos}2\phi\right)\mathrm{sin}\theta+\left(\hat{Q}^{yz}\mathrm{cos}\phi-\hat{Q}^{xz}\mathrm{sin}\phi\right)\mathrm{cos}\theta\;,
\en
where $\theta$ is the pole angle and $\phi$ is the azimuthal angle in commonly used polar coordinate system, and~\cite{Varshalovich:1988ye}
\eq
(\hat{Q}^{ik})_{lm} =(\hat{Q}_{ik})_{lm} & = & -\frac{1}{2}\left(\delta_{il}\delta_{km}+\delta_{im}\delta_{kl}-\frac{2}{3}\delta_{ik}\delta_{lm}\right)\;,\quad\left(i,k,l,m=x,y,z\right)\;.
\en

\bibliography{ref_v2}

\end{document}